\documentclass[sigconf, table,xcdraw]{acmart}

\usepackage{multirow}
\usepackage{enumitem}
\usepackage{balance}

\AtBeginDocument{%
  }

\copyrightyear{2025}
\acmYear{2025}
\setcopyright{acmlicensed}\acmConference[RecSys '25]{Proceedings of the Nineteenth ACM Conference on Recommender Systems}{September 22--26, 2025}{Prague, Czech Republic}
\acmBooktitle{Proceedings of the Nineteenth ACM Conference on Recommender Systems (RecSys '25), September 22--26, 2025, Prague, Czech Republic}
\acmDOI{10.1145/3705328.3759331}
\acmISBN{979-8-4007-1364-4/2025/09}


\begin{abstract}

In modern recommender systems, experimental settings typically include filtering out cold users and items based on a minimum interaction threshold. However, these thresholds are often chosen arbitrarily and vary widely across studies, leading to inconsistencies that can significantly affect the comparability and reliability of evaluation results. In this paper, we systematically explore the cold-start boundary by examining the criteria used to determine whether a user or an item should be considered cold. Our experiments incrementally vary the number of interactions for different items during training, and gradually update the length of user interaction histories during inference. We investigate the thresholds across several widely used datasets, commonly represented in recent papers from top-tier conferences, and on multiple established recommender baselines. Our findings show that inconsistent selection of cold-start thresholds can either result in the unnecessary removal of valuable data or lead to the misclassification of cold instances as warm, introducing more noise into the system.

\end{abstract}

\makeatletter
\@printpermissionfalse
\makeatother

\begin{document}

\title{Recommendation Is a Dish Better Served Warm}

\author{Danil Gusak}
\orcid{0009-0008-1238-6533}
\affiliation{%
  \institution{AIRI, Skoltech}
  \city{Moscow}
  \country{Russian Federation}
}
\email{danil.gusak@skoltech.ru}
\authornote{Authors contributed equally to the paper}

\author{Nikita Sukhorukov}
\orcid{0009-0005-0965-4654}
\affiliation{%
  \institution{AIRI, Skoltech}
  \city{Moscow}
  \country{Russian Federation}
}
\email{sukhorukov@airi.net}
\authornotemark[1]

\author{Evgeny Frolov}
\orcid{0000-0003-3679-5311}
\affiliation{%
  \institution{AIRI, HSE University, Skoltech}
  \city{Moscow}
  \country{Russian Federation}
}
\email{frolov@airi.net}

\maketitle

\section{Introduction}

Recommender systems (RS) are designed to address the challenge of matching users with relevant items from a vast catalog of options. A common characteristic across various datasets is the skewed data distribution, often characterized by a ``\emph{short head}'' of highly popular items and a ``\emph{long tail}'' of niche or less frequently consumed items \cite{park2008longtail, wu2008longtail}. While the short head accounts for a significant portion of total interactions due to its popularity, the long tail forms most of the item space, offering opportunities for personalized and diverse recommendations. The primary objective of a recommender system is to discover the most relevant items from this sparse long tail, producing non-obvious suggestions beyond simple popular recommendations and enhancing user experience.

\begin{figure}[t!]

    \centering
\setlength{\abovecaptionskip}{6pt}
\setlength{\belowcaptionskip}{0pt} 
\includegraphics[width=0.98\columnwidth]{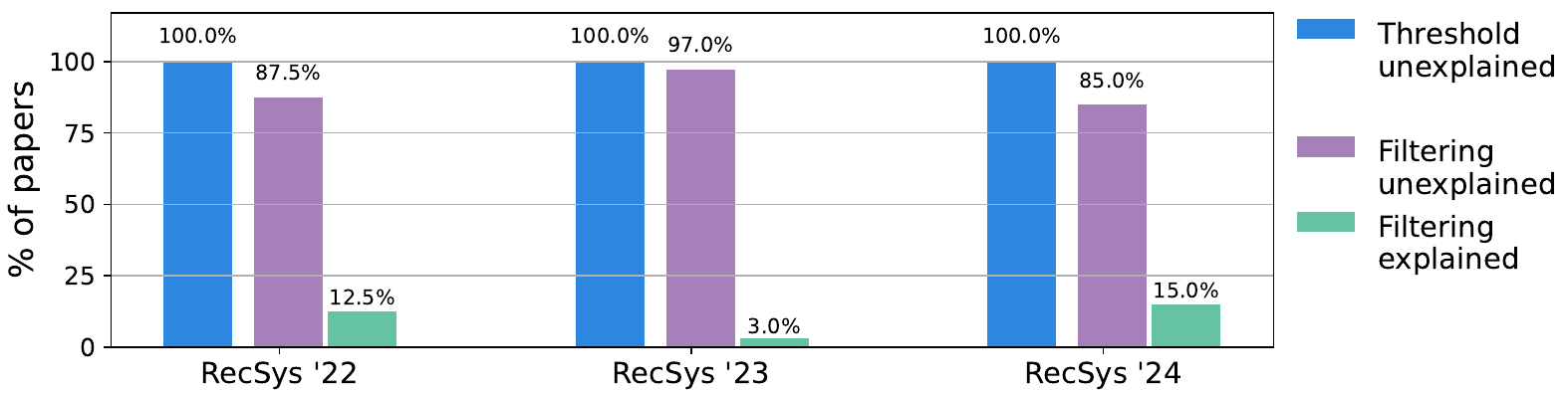}
    \caption{Trends in threshold and filtering justification. None of the papers that mentioned filtering provided an explanation for the chosen threshold value. Only 3--15\% of papers provided an explanation for applying filtering.}
\label{fig:paper_stats}
\end{figure}

However, long-tail items often suffer from insufficient interaction data, introducing many challenges. Items with too few interactions can induce unnecessary noise into the collaborative filtering signal, degrading model quality \cite{xue2019deepcf, fan2023graphcollabdenoising, wang2021denoisingimplicitfeedbackforrecs}.
Hence, determining when interaction history is ``warm'' enough, i.e., containing sufficient feedback for reliably extracting useful behavioral patterns, becomes a crucial task. Additionally, the sheer size of the long tail can overwhelm the model, heavily increasing computational complexity during both training and inference \cite{mezentsev2024scalable, gusak2024rece}. Entities (users or items) with very limited interaction histories are closely tied to the cold-start problem~\cite{coldstart, sukhorukov2025maximum}, where accurate recommendations require additional information, such as content-based features or metadata \cite{singh2024bettergeneralizationsemanticids, frolov2019hybridsvd, sukhorukov2025maximum}.

Identifying users or items with insufficient interaction data is critical for addressing the cold-start problem. Entities below a certain threshold of interaction activity are generally considered ``cold'', meaning they lack sufficient historical data to generate meaningful recommendations using traditional collaborative filtering techniques. However, the range of adequate values of this threshold has not been properly investigated. The threshold is often set to some default based on anecdotal evidence or conventional heuristics, which may undermine its core purpose of ensuring input data consistency and reliability.

To assess the prevalence of dataset filtering practices, we analyzed all papers from the RecSys'22--24 conference proceedings. Out of sixty-five papers that mentioned filtering, none offered an explanation for their chosen threshold, and only six works (9\%) provided justification for applying filtering, citing issues such as data sparsity and coverage. The remaining papers (91\%) did not explain the rationale behind their filtering choices, highlighting the lack of systematic reasoning in filtering decisions within the RS community (Figure~\ref{fig:paper_stats}). Therefore, establishing this threshold is essential for differentiating between cold-start and standard recommendation scenarios and generating recommendations that accurately mirror user behavior.

Motivated by these challenges, we propose a model‑agnostic technique that allows us to identify two distinct ``cold-warm'' thresholds: one for items and another for users. These thresholds serve as critical indicators for determining whether an entity should be treated as cold or part of the standard scenario \cite{marlin2004standardscenario}. 

For \emph{items}, our approach focuses on identifying the minimum number of interactions required to treat an item as warm. This is achieved by systematically varying the number of interactions each item has in the training set. We analyze the performance of recommendation algorithms under different interaction thresholds to evaluate the point at which an item transitions from being considered "cold" to contributing meaningful collaborative signals.

Similarly, for \emph{users}, we aim to determine the minimum length of interaction history needed to produce sensible and accurate recommendations. To achieve this, we analyze how recommendation quality varies with the length of a user's interaction history. By varying this length, we assess the impact on recommendation accuracy, diversity, and relevance. The resulting threshold provides a clear guideline for distinguishing between users who require auxiliary information (e.g., content-based features) due to insufficient interaction data and those who can benefit from traditional collaborative filtering techniques.

In summary, our main contributions are:
\begin{itemize}
[leftmargin=*,label=\textbullet,nosep]
    \item We present a training scheme to determine the item-based interaction threshold;
    \item We propose an inference scheme to determine the user-based interaction threshold;
    \item We conduct systematic experiments across multiple datasets and recommender models, demonstrating that ignoring threshold variations can lead to suboptimal or inconsistent evaluation.

\end{itemize}
\section{Related Work}
\label{sec:relatedwork}

Most collaborative filtering studies apply thresholds to exclude cold users or items, commonly using $p$-core filtering \cite{sun2021areweevaluatingrigorouslypcore, sachdeva2020howusefularereviews, dacrema2019arewereallymakingmuchprogress}. This method iteratively removes users with fewer than $p$ interactions and items with fewer than $p$ users; some variants apply this thresholding only to users or items~\cite{hidasi2023widespread, mezentsev2024scalable}. While this reduces sparsity and enhances data density, the selection of $p$ is typically heuristic and lacks systematic justification \cite{hidasi2023widespread, sun2021areweevaluatingrigorouslypcore}.
\citet{visnovsky2021cold} addressed the cold-start problem by estimating the minimal number of interactions required for a user to be considered non-cold. Authors clustered users into groups of approximately $k$ users using the K-means algorithm, and determined the optimal interaction threshold using the Davies-Bouldin index \cite{Davies1979ACS} as a clustering quality measure. However, the experiments were limited to two TV datasets.

Overall, the topic of selecting a cold-start threshold remains almost unexplored in the literature~\cite{handbook}.
In this work, we propose a straightforward approach for identifying the minimal thresholds for classifying instances as cold. Our method accounts for both user- and item-based settings, covering the main RS formulations. Unlike previous approaches that rely on empirical tuning or domain-specific assumptions, our technique dynamically adapts to dataset characteristics, ensuring robustness across diverse domains. By explicitly defining these thresholds, we aim to provide better insights into the cold-start problem and enable more tailored RS solutions.

\section{Methods}
\label{sec:method}

We investigate the cold-start thresholds for both items and users. Our goal is to determine how many interactions are required for standard collaborative and sequential recommendation models to begin producing meaningful results. We fix the training-validation-test splits and vary the number of interactions in the histories of the items or users under study. Detailed information on the experimental setup can be found in Section \ref{sec:eval}.
In this paper, we focus on the established next-item prediction (NIP) task, recommending items to users. Although it is possible to study the reverse setting of predicting users for items, our focus is on the setup with the natural pattern of consumption.

\begin{table}[t]

\setlength{\abovecaptionskip}{3pt}
\caption{\textbf{{Statistics of the datasets}}} \label{tab:datasetStats}
\resizebox{1\columnwidth}{!}{%
\begin{tabular}{lrrrrrr}
\bottomrule
\multirow{2}{*}{Dataset} &
  \multirow{2}{*}{Users} &
  \multirow{2}{*}{Items} &
  \multirow{2}{*}{Interact.} &
  \multirow{2}{*}{Density} &
  \multirow{2}{*}{\begin{tabular}[c]{@{}r@{}}Avg. User\\ Interact.\end{tabular}} &
  \multirow{2}{*}{\begin{tabular}[c]{@{}r@{}}Avg. Item\\ Interact.\end{tabular}} \\
                                    &        &        &           &        &      &      \\ \toprule
ML-1M \cite{ml1}                    & 6,040  & 3,706  & 1,000,209 & 4.47\% & 166  & 270  \\
BeerAdvocate \cite{mcauley2012beer} & 7,606  & 22,307 & 1,409,494 & 0.83\% & 185  & 63.2 \\
Behance \cite{behance}              & 8,097  & 32,434 & 546,284   & 0.21\% & 67.4 & 16.8 \\
Beauty \cite{mcauley2015imagebased} & 26,116 & 17,037 & 230,481   & 0.05\% & 8.83 & 13.5 \\ \bottomrule
\end{tabular}%
}
\end{table}

\subsection{Item-Based Threshold} \label{sec:cold_items}

For an item-level analysis, we investigate how the number of training interactions for an item affects its likelihood of being recommended.
For each item $i \in I$, where $I$ is the dataset item catalog, let $M_{(i)}$ be the set of its training interactions across all users. To evaluate how the number of interactions impacts recommendations, we consider a subsample of $M_{(i)}$ of size $N$ (with $1 \leq N \leq |M_{(i)}|$). For item $i$, we randomly select $N$ interactions from $M_{(i)}$ and construct a modified training set, $\tau_{N}$, removing the remaining $|M_{(i)}|- N$ interactions of the investigated item. We then retrain our recommendation model using $\tau_{N}$ and measure the ranking performance for item $i$ in the test set.
\emph{By varying $N$ and retraining the model each time, we can track the dynamics of positional rank of item $i$ in the test recommendations}. Note that we sample $N$ random interactions for an item instead of using the latest $N$, because it could bias results by overrepresenting items with recent interactions, affected by shifting popularity and user preference drift \cite{sun2023take, ji2023critical}.
We conduct the same procedure for multiple items from $I$, and average results for each $N$ to reveal overall trends.

\begin{figure*}[h!t!]
    \centering
\setlength{\abovecaptionskip}{6pt} 
    \setlength{\belowcaptionskip}{0pt} 
\includegraphics[width=1.0\textwidth]{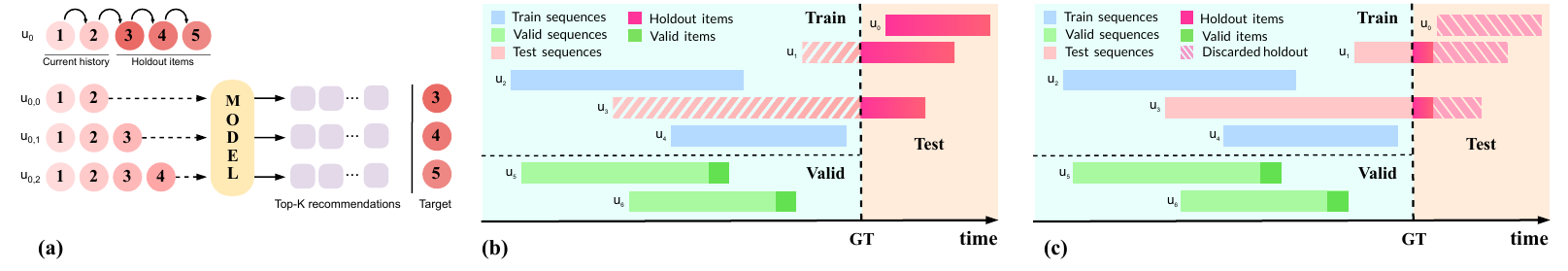}
    \caption{(a) Successive evaluation scheme. (b-c) Data splitting setups for \emph{Item-Based} setting (b) and  \emph{User-Based} setting (c).}
\label{fig:setups}
\end{figure*}

\subsection{User-Based Threshold} \label{sec:cold_users}

For a user-level analysis, we \textit{fix the trained model and, during the inference phase, investigate how the number of interactions available for a test user influences the user's recommendation quality} in terms of standard RS metrics. Let $U_t$ denote the set of test users and, for each test user $u_t \in U_t$, let $L_{(u_t)}$ represent their complete interaction history. For a given number $N$ (with $1 \leq N \leq |L_{(u_t)}|$), we uniformly sample $N$ interactions from $L_{(u_t)}$ to create a truncated history, denoted by $L^N_{(u_t)}$, and remove the remaining interactions. Uniform sampling allows to preserve the original popularity distribution, since popular items are more likely to appear in users' histories and be sampled. The truncated history is then provided as input to a trained recommender model to predict the holdout interactions and compute metrics. By randomizing, we avoid bias toward either very old or very recent interactions and capture an average effect.
Although the ordering of the sampled interactions is not critical for standard collaborative models, it is essential for sequential models (e.g., SASRec \cite{kang2018self}).
To address this, we sort sampled interactions $L^N_{(u_t)}$ by timestamp, ensuring that the natural consumption sequence is preserved and that a sequential model captures at least the user's global intent.
The procedure is repeated across all test users in $U_t$, and the resulting metrics are averaged for each value of $N$.

\section{Experiments}
\label{sec:expers}

\subsection{Experimental Settings}
\label{sec:expsetup}

In this section, we reveal details of the settings and discuss the results. The code for reproduction is available in the repository\footnote{\url{https://github.com/dalibra/recommendation-dish-better-served-warm}\label{github}}.


\subsubsection{Datasets}

We conduct our experiments on four popular real-world datasets:
MovieLens-1M \textbf{(ML-1M)} \cite{ml1}, 
\textbf{BeerAdvocate} \cite{mcauley2012beer},
\textbf{Behance} \cite{behance},
and \textbf{Amazon Beauty} \cite{mcauley2015imagebased}.
The datasets vary significantly in domains, density, and average number of interactions per user and item, allowing us to evaluate
approaches under various conditions. The statistics of the datasets are summarized in Table \ref{tab:datasetStats}.

\subsubsection{Evaluation Setup}
\label{sec:eval}

Data splitting using the leave-one-out approach, selecting the last interaction of each user, is common in previous studies \cite{2024autoregressive, kang2018self}. 
However, it leads to data leakage, compromising the reliability of the results \cite{timetosplit2025, sun2023take, ji2023critical}.
To address this issue in the NIP
setting, we set a \emph{global timepoint} (GT) at the $0.9$-quantile of all interactions and take preceding interactions into the train and validation subsets, splitting them randomly by the users~\cite{timetosplit2025}. For validation, we treat each user’s last interaction as the ground truth while using all earlier interactions as the input sequence; this set is employed for hyperparameter tuning ($20$ random points from the hyperparameter grid) and early stopping. All users with interactions after the global timepoint (holdout interactions) are considered test users, and their earlier interactions are treated as input sequences~\cite{timetosplit2025}. Next, we explore the specific aspects of the evaluation that are directly tied to the Item-based and User-based setups.
\paragraph{Item-Based}\label{sec:metr_items}

In this setup, we remove interactions of the investigated item $i$ from both the test and validation to eliminate any specific effects related to this item. We also exclude the interactions of test users from the training, ensuring that the train data remains consistent across different items -- retaining these users would result in different interaction counts per test user depending on the item $i$, as the test filtering would vary with $i$. Overall, we carefully design the setup so that the training set remains mostly unchanged, allowing for a variation of up to $\lvert M_i \rvert - \text{max}(N)$ interactions, and is fully separated from the test and validation sets, minimizing any bias in the experiment (Figure \ref{fig:setups}b). Finally, \emph{recommendation stability}~\cite{olaleke2021dynamic} stays within a 1\% range for all $N$, 
\textit{confirming that the sampling strategy is robust and preserves semantics of the datasets}.

We adopt \emph{successive evaluation} scheme \cite{timetosplit2025, sun2023take}, as shown in Figure~\ref{fig:setups}a. In this approach, a user's current interactions are incrementally expanded by adding holdout interactions one at a time. After each step, we generate a top-K recommendation list and compare it against the next holdout interaction. As a result, for $n$ holdout interactions, we obtain $n$ separate metric evaluations, cheaply enhancing robustness while effectively capturing changing user behavior~\cite{timetosplit2025, hidasi2023widespread}.

Given the large size of item catalogs in recommender system datasets, retraining the model for every item and for each value of \(N\) would be computationally prohibitive -- the total number of retraining iterations for a single dataset would be up to $|\{N_j\}| \times | I | \times \text{(\# models)}$.
Therefore, to construct sufficient confidence intervals (CI) without excessive computational overhead, we perform this procedure on a subset of $S$ uniformly sampled items from the training catalog ($S$ = 1000 in our experiments). Successive evaluation further improves the robustness of results.

\paragraph{User-Based}\label{sec:metr_users}

In the Item-based setting, successive evaluation is suitable since we focus solely on whether a certain item appears in recommendation lists. In contrast, in the User-based setup, we compute metrics against the ground truth, which introduces degradation effects as model predictions move further from the GT~\cite{sun2023take, ji2023critical}. To mitigate this, we retain only the first interaction of each test user after the GT as the ground truth and discard the rest (Figure~\ref{fig:setups}c).

For User-based analysis, \textit{we fix a single trained model for inference} (Section \ref{sec:cold_users}), \textit{avoiding costly retrainings} required in the Item-based setup. This allows for cost-efficient use of the whole set of test users as investigated users for the analysis and final metric computation.

\subsubsection{Metrics}
\label{sec:metrics}

\paragraph{Item-Based}\label{sec:meth_items}

We study how often an investigated item appears in top-K recommendations, varying its history length $N$.
Following Section \ref{sec:cold_items}, we extract the positional rank of the item from the recommendation lists provided to test users. Instead of computing the standard metrics against the ground truth item, we define \emph{modified HitRate}$^{*}$ (HR$^{*}$@K) and \emph{modified Normalized Discounted Cumulative Gain}$^{*}$ (NDCG$^{*}$@K), which specifically measure the presence frequency and ranking of the investigated item:

\begin{equation}
\text{HR*@K(i)} = \frac{1}{|U|}  \sum_{u \in U} \mathbb{I} [ i \in R^K_u ],
\end{equation}

\begin{equation}
\text{NDCG*@K(i)} = \frac{1}{|U|} \sum_{u \in U} \frac{\mathbb{I}[i \in R_u^K]}{\log_2(\text{rank}_u(i) + 1)},
\end{equation}
where $I$ is the dataset item catalog, $U$ is the evaluation set of users, $R^K_u$ is the top-K recommendation list for user $u$, and $\text{rank}_u(i)$ is the rank of item $i$ in $R^K_u$.

We compute these metrics at K = 1, 5, 10, 50, 100, and average the results over items with at least $N$ interactions in the training set ($I_N$):

\begin{equation}
\text{HR*@K} = \frac{1}{|I_N|}  \sum_{i \in I_N} \text{HR*@K(i)},
\end{equation}
\begin{equation}
\text{NDCG*@K} = \frac{1}{|I_N|} \sum_{i \in I_N} \text{NDCG*@K(i)}.
\end{equation}

\paragraph{User-Based}\label{sec:meth_users}
In this setup, we compute \textit{standard} top-K recommendation metrics.
Following best practices \cite{ Dallmann_2021, 50163}, we use unsampled NDCG@10 and HR@10 to evaluate performance, and average results across all users with at least N interactions in the test set.

\subsubsection{Models}
\label{sec:implementation_details}

We conduct experiments with four established recommender system models: \textbf{EASE}$^{\text{\textbf{R}}}$ \cite{steck2019embarrassingly}, an effective linear model that captures item co-occurrence through a closed-form solution; \textbf{PureSVD} \cite{Cremonesi2010PerformanceOR}, which uses singular value decomposition to uncover latent preferences; \textbf{ItemKNN} \cite{itemknn}, a neighborhood-based approach that relies on item similarities computed from collaborative interactions; and sequential model \textbf{SASRec-CE} \cite{Klenitskiy_2023}, a variant of the original SASRec \cite{kang2018self} that employs full cross-entropy loss to enable state-of-the-art performance \cite{Klenitskiy_2023}.

\subsection{Results}
\label{sec:results}

Since interaction counts vary across both users and items (Table~\ref{tab:datasetStats}), some of them may have fewer interactions than a given $N$. Consequently, the number of users and items available for evaluation drops as $N$ increases, leading to wider confidence intervals at higher $N$ values. Our experiments also showed that the results remain consistent regardless of whether the \emph{filter seen} \cite{filter_seen} step is applied.

\paragraph{Item-Based}

\begin{table}[b]
\setlength{\abovecaptionskip}{6pt}
\caption{Estimated "cold-warm" regime thresholds for users and items on different datasets} \label{tab:results}
\resizebox{1\columnwidth}{!}{%
\begin{tabular}{llcccc}
\toprule
\textbf{Setup}                       & \textbf{Model}             & \textbf{ML-1M} & \textbf{BeerAdvocate} & \textbf{Behance} & \textbf{Amazon Beauty} \\ \midrule
\multirow{4}{*}{Item-Based} & SASRec-CE         &   9\footnotesize{$\pm$1.1}   &      10\footnotesize{$\pm$1.5}     &   15\footnotesize{$\pm$1.7}    &     10\footnotesize{$\pm$2.1}      \\
                            & ItemKNN           &   9\footnotesize{$\pm$1.8}   &      6\footnotesize{$\pm$1.2}      &   12\footnotesize{$\pm$1.3}    &     6\footnotesize{$\pm$1.1}        \\
                            & PureSVD           &   9\footnotesize{$\pm$1.4}   &      6\footnotesize{$\pm$1.2}      &   6\footnotesize{$\pm$1.0}     &     6\footnotesize{$\pm$0.9}       \\
                            & EASE$^{\text{R}}$ &   10\footnotesize{$\pm$1.0}   &      8\footnotesize{$\pm$1.1}     &   6\footnotesize{$\pm$1.1}     &     8\footnotesize{$\pm$1.1}       \\ \midrule
\multirow{4}{*}{User-Based} & SASRec-CE         &   6\footnotesize{$\pm$0.8}   &     10\footnotesize{$\pm$2.1}       &   10\footnotesize{$\pm$1.9}    &     12\footnotesize{$\pm$1.8}         \\
                            & ItemKNN           &   4\footnotesize{$\pm$1.4}   &     8\footnotesize{$\pm$1.3}        &   10\footnotesize{$\pm$2.3}    &     4\footnotesize{$\pm$1.2}         \\
                            & PureSVD           &   4\footnotesize{$\pm$1.3}   &     8\footnotesize{$\pm$1.2}        &   6\footnotesize{$\pm$1.4}     &     4\footnotesize{$\pm$1.1}         \\
                            & EASE$^{\text{R}}$ &   5\footnotesize{$\pm$1.1}   &     8\footnotesize{$\pm$1.6}        &   10\footnotesize{$\pm$1.7}    &     4\footnotesize{$\pm$1.1}         \\ \bottomrule
\end{tabular}%
}
\end{table}

By the experimental design, we remove some interactions of the investigated items. It is worth noting that such a setup disrupts users' natural sequence, which may introduce bias by breaking native transitions or creating new, unintended ones. However, our experiments show that removing a small number of interactions does not substantially affect the results (Section~\ref{sec:metr_items}).

\begin{figure}[t]
    \centering
\setlength{\abovecaptionskip}{3pt} 
    \setlength{\belowcaptionskip}{0pt} 
\includegraphics[width=1\columnwidth]{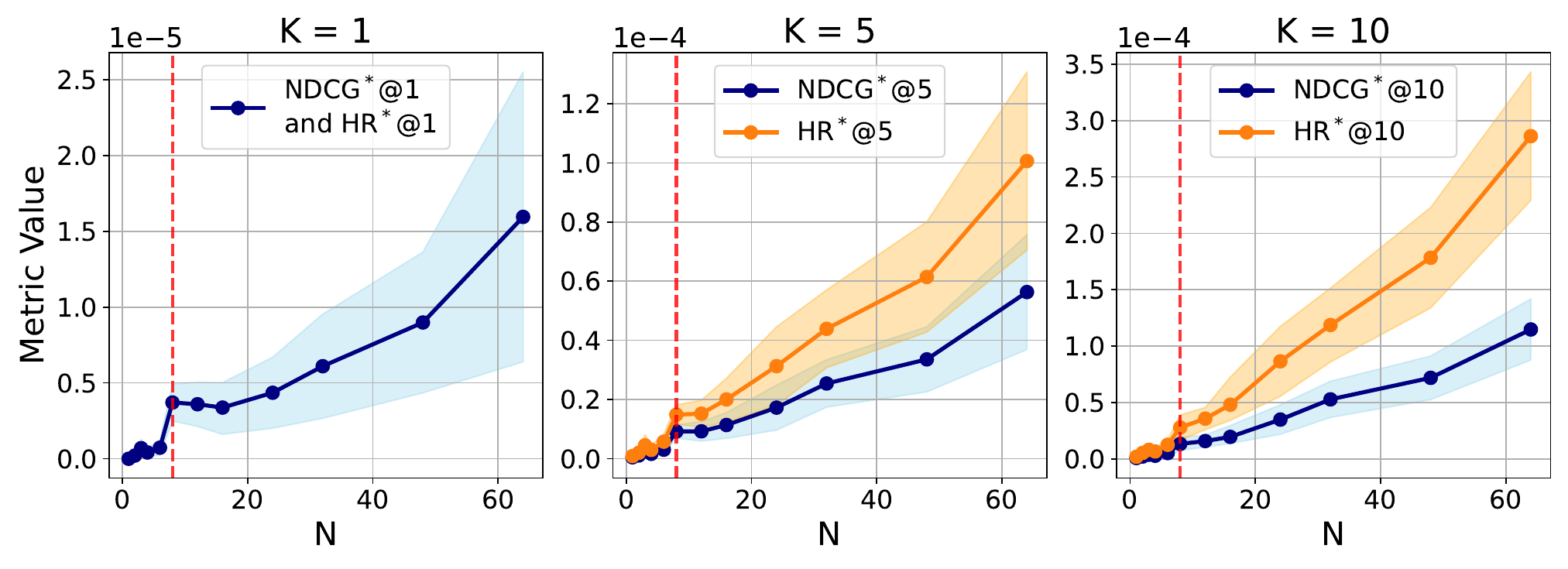}
    \caption{Dynamics of metrics (95\% CI) for different @K for ML-1M dataset and SASRec-CE in the \emph{Item-Based} setup.}
\label{fig:ml1_sasrec_items}
\end{figure}

To determine numerical thresholds that distinguish cold items from warm ones, we analyze how modified metrics (HR$^*$ and NDCG$^*$) vary with the frequency of investigated item interactions in the training data. A sharp increase in performance (Figure~\ref{fig:ml1_sasrec_items}, vertical red lines) indicates the cold-to-warm transition, reflecting the point at which \textit{an item accumulates sufficient interaction data to be effectively incorporated into the model’s learned representation}.
To identify the transition point, which we refer to as ``performance shift'' from cold to warm, we apply a popular sliding window approach~\cite{sliding_window1, sliding_window2}. 
For each window applied along $N$, we compute the slope of the best-fit line by solving a least squares regression problem. The interaction count corresponding to the window with the steepest positive slope marks the threshold, indicating the point where metrics begin to improve rapidly with increased item frequency. 

The resulting threshold values, summarized in Table~\ref{tab:results}, are stable across models for ML-1M; however, for other datasets, thresholds vary notably, ranging from 6 to 15 interactions per item.
The metrics dynamics for ML-1M and the SASRec-CE model are illustrated in Figure~\ref{fig:ml1_sasrec_items}. As K increases, the recommendation lists are more likely to include the investigated items, leading to smoother metric curves with reduced variance. However, this results in a smoother ``performance shift'', making the transition point less distinct.

\paragraph{User-Based}

\begin{figure}[t]
\setlength{\abovecaptionskip}{3pt}
\setlength{\belowcaptionskip}{0pt} 
    \centering
    \includegraphics[width=0.99\columnwidth]{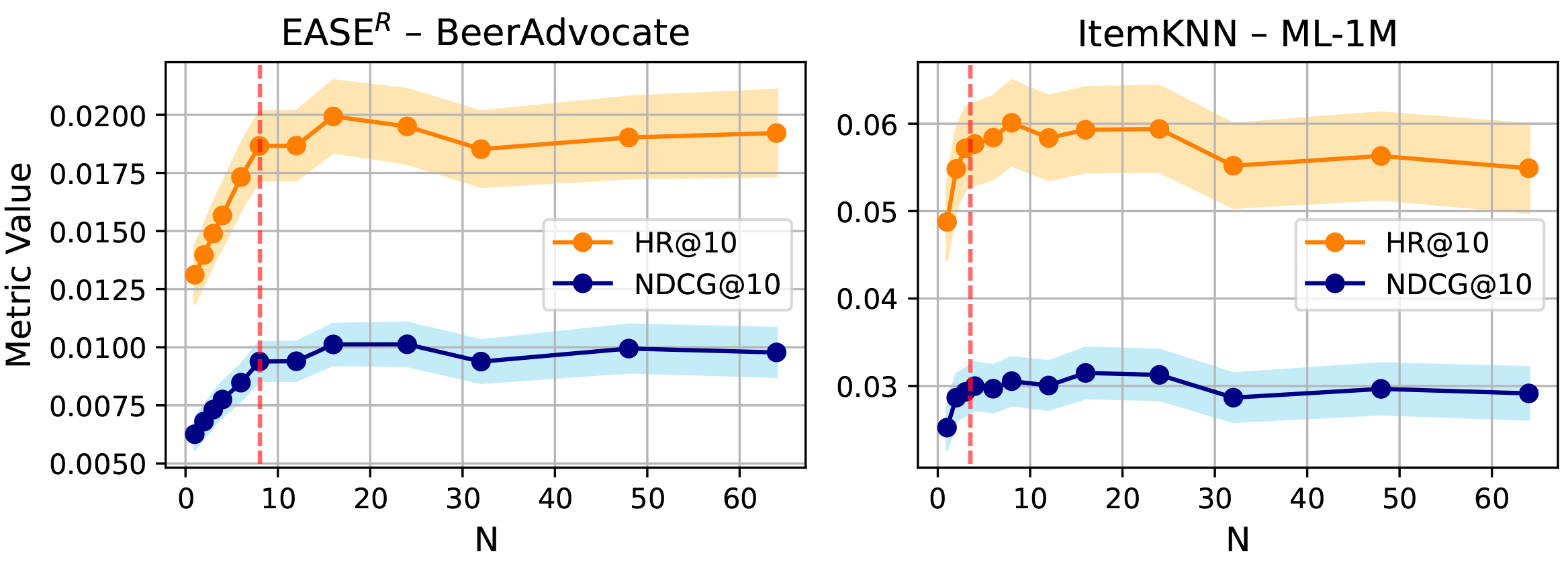}
    \caption{Dynamics of metrics (95\% CI) with respect to number of interactions $N$ in user's history in the \emph{User-Based} setup.}
    \label{fig:cold_users}
\end{figure}

In this setup, the identified thresholds mark the point of quality saturation (Figure~\ref{fig:cold_users}).
We applied the same method used in the Item-based analysis, identifying the threshold point through slopes of local linear regressions. At this point, the likelihood of capturing all essential items increases significantly, and further interactions provide only minor performance gains. However, excessively long interaction histories (large $N$) \textit{can dilute the most informative signals} \cite{cremonesi2012profiling, lathia2010temporaldiversity}, leading to performance stagnation. This suggests that beyond a certain amount of interactions, adding new data introduces noise rather than useful information.

Table \ref{tab:results} outlines the thresholds for the User-based setup. For classical collaborative filtering models (EASE$^{\text{R}}$, PureSVD, and ItemKNN), thresholds are similar within a dataset, suggesting that, despite differing underlying mechanisms, these models require comparable interaction data volumes to perform effectively.
Notably, the sequential SASRec-CE model consistently requires more interactions than classical non-sequential models before exhibiting a behavior shift. This suggests that SASRec-CE needs longer user histories to capture consumption patterns optimally. Furthermore, given substantial variability in thresholds across datasets, researchers should \textit{carefully select appropriate values} to ensure consistent evaluation and optimal performance. When thresholds vary between models, adopting the maximum threshold ensures a fair and consistent evaluation setup across all models.

\section{Conclusion}
\label{sec:conclusion}

In this study, we explored the ``cold-warm'' transition problem in recommender systems by identifying minimal interaction thresholds for both items and users. Our findings show that arbitrary thresholding can lead to data loss or misclassification of cold entities, which in turn biases evaluation results. To address this problem, we introduced a robust, model-agnostic approach.
For items, we varied the number of training interactions and applied successive evaluation. For users, we kept the model fixed and sampled different lengths of their interaction history at inference time.

Our findings show that proper threshold selection has a meaningful impact on observed performance. In the user-based setup, performance tends to plateau beyond a certain amount of considered interactions, whereas in the item-based scenario, we observe sharp improvements in evaluation metrics, indicating a cold-to-warm transition.
These insights support the use of reasonable data-driven thresholding of interactions instead of relying on arbitrary filtering methods.
With this study, we also hope to highlight new research opportunities. In addition to improving the scalability of the method and extending its application beyond the collaborative filtering setups, one could also utilize the identified thresholds to train and inference RS models using a split-representation approach, where cold users or items are marked with a special token rather than being simply filtered out. This would enable collaborative signal propagation through shared tokens, helping to mitigate the dataset filtering issues.

\bibliographystyle{ACM-Reference-Format}
\balance
\bibliography{sections/bibliography}

\end{document}